# Magnetically Multiplexed Heating of Single Domain Nanoparticles

M. G. Christiansen,[1)] R. Chen,[1)] and P. Anikeeva[1,a)]

[1]Department of Materials Science and Engineering, Massachusetts Institute of Technology, Cambridge, Massachusetts, 02139, USA

**Abstract:**

Selective hysteretic heating of multiple collocated sets of single domain magnetic nanoparticles (SDMNPs) by alternating magnetic fields (AMFs) may offer a useful tool for biomedical applications. The possibility of "magnetothermal multiplexing" has not yet been realized, in part due to prevalent use of linear response theory to model SDMNP heating in AMFs. Predictive successes of dynamic hysteresis (DH), a more generalized model for heat dissipation by SDMNPs, are observed experimentally with detailed calorimetry measurements performed at varied AMF amplitudes and frequencies. The DH model suggests that specific driving conditions play an underappreciated role in determining optimal material selection strategies for high heat dissipation. Motivated by this observation, magnetothermal multiplexing is theoretically predicted and empirically demonstrated for the first time by selecting SDMNPs with properties that suggest optimal hysteretic heat dissipation at dissimilar AMF driving conditions. This form of multiplexing could effectively create multiple channels for minimally invasive biological signaling applications.

**Text:**

Magnetic fields provide a convenient form of noninvasive electronically driven stimulus that can reach deep into the body because of the weak magnetic properties and low conductivity of tissue. Single domain magnetic nanoparticles (SDMNPs) may act as transducers of heat for such remote stimulation in the presence of an alternating magnetic field (AMF).[1, 2] Despite decades of research, biological applications of magnetically induced heat dissipation remain limited to remote activation of individual processes such as controlled cell death[3] or, more recently, initiation of a single biochemical pathway.[4, 5] Many biomedical applications would benefit from the capability to independently and remotely control multiple pathways or cell types in close spatial proximity through selectively heating particle sets with differing magnetic properties by varying the driving conditions of the AMF.

The possibility for such multiplexed magnetic heating has not yet been demonstrated, in part because it relies upon recognizing the influence of the applied AMF on stochastic magnetization reversal. Though such

---

[a)] **Author to whom correspondence should be addressed. Electronic mail: anikeeva@mit.edu.**

dependence has long been studied,[6, 7] it is frequently overlooked in the context of SDMNP heat dissipation due to prevalent use of linear response theory (LRT),[8] which neglects the effect of the applied AMF on the effective barrier to thermally assisted reversal.[9] Despite attempts to repair LRT by adding field dependence to the relaxation time,[10] this simplified theory is unable to predict the non-elliptical shapes of experimentally observed hysteresis at high AMF amplitudes,[11] indicating the need for more general approaches. One alternative is to model coherent precession and reversal with the Landau–Lifshitz–Gilbert equation, incorporating a fictitious stochastically varying thermal field in addition to the effective field.[12, 13] Dynamic hysteresis (DH) models offer a simpler alternative to describe nonlinear effects, treating the coherent reversal of particle moments as kinetically limited by thermal activation over a time varying energy barrier determined by both the SDMNP anisotropy and the Zeeman energy of its moment in the field (See Fig. 1(a)).[14, 15] In this manuscript, we compare the predictions of a DH model with experimental calorimetry data collected using high quality magnetic nanomaterial sets at varied AMF amplitudes and frequencies. We are then able to demonstrate the possibility of magnetothermal multiplexing with specific loss powers by separately heating two sizes of iron oxide ($Fe_3O_4$) and manganese ferrite ($Mn_xFe_{2-x}O_4$) SDMNPs with AMFs of differing amplitude and frequency.

In order to represent the essential physics of DH, it is useful to introduce the graphical concept of $\sigma$-$\xi$ space, where $\sigma$ and $\xi$ are unitless parameters defined by the scale of the anisotropy energy of a SDMNP and the Zeeman energy of its moment in an applied field normalized with respect to the thermal energy, respectively.

$$\sigma \equiv \frac{KV}{k_B T}$$

$$\xi \equiv \frac{\mu_0 H_0 M_s V}{k_B T}$$

Here, $K$ is the effective anisotropy of the SDMNP, $V$ is the volume, $H_0$ is the field amplitude, $M_s$ is the saturation magnetization, and $\mu_0$ is the permeability of free space. These definitions of $\sigma$ and $\xi$ assume idealized spherical single crystal SDMNPs with uniform magnetization and effective anisotropy that scales with volume regardless of contributions from surface, shape, or magnetostrictive effects. While these assumptions do not fully describe the complexity of real SDMNPs,[16] identifying the physical origins of $\sigma$ is not conceptually essential for multiplexing. We assume the SDMNPs to be uniaxial with an estimated $K_{U1} \approx |K_1|$ ($K_{U1}$ is the anisotropy energy associated with the first order term for uniaxial anisotropy and $K_1$ is analogous for cubic anisotropy), rotationally fixed, and non-interacting such that the values of $\sigma$ and $\xi$ set the effective barrier to reversal and, together with the timescale set by the frequency of the AMF, fully determine the hysteresis behavior of the system in the model (Fig. 1(a)). Each point in $\sigma$-$\xi$ space corresponds to a parametrically defined hysteresis loop with an area proportional to the energy dissipated per SDMNP, per cycle of the AMF (Fig. 1). The surface plotted in Fig. 1(b) summarizes areas of hysteresis loops (Fig. 1(c)) corresponding to different values of $\sigma$ and $\xi$.

Defining the anisotropy field $H_k$ as the applied field at which the barrier to reversal vanishes, the ratio $\xi/(2\sigma)$ can be rewritten as $H_0/H_k$, describing a linear path through $\sigma$-$\xi$ space through the origin with a slope determined by the magnitude of the AMF amplitude (Figs. 1(b) and 1(c)). In the limit of low AMF amplitudes (i.e., $H_0 \ll H_k$), a critical $\sigma_{crit}$ value occurs where the mean lifetime of stochastic magnetization reversal in the absence of an applied AMF is equal to the timescale of measurement set by its frequency.[12, 14] For low AMF amplitudes, $\sigma_{crit}$ corresponds to the maximum predicted by LRT. Moments on either side of this maximum either escape more rapidly or more slowly than the timescale set by the frequency of the field, which defines the superparamagnetic and ferromagnetic regimes, respectively. DH simulations predict markedly different dependence of loss power on AMF amplitude for the two regimes. (See Figs. 1(c)-1(e) for plots of the hysteresis loops.)

With this graphical representation, the strategy for efficiently producing maximal loss powers with the constraint of a given $H_0 f$ product seems clear (Fig. 1(c)). Large SDMNPs driven by AMFs with amplitudes approaching or exceeding $H_k$ at high frequencies will produce hysteresis loops approaching the theoretical limit set by the material, provided that the fastest available relaxation process is coherent reversal[17] and the detailed dynamical picture of damped precession[12] can be neglected. Increasing $H_0$ beyond $H_k$ does increase the effective coercive field and loop area, but with diminishing marginal returns rather than the linear increase predicted by LRT. This strategy was borne out experimentally in the highest heating rates reported, where a high $H_0 f$ product is used and "tuning" of materials parameters has little to do with attaining $\sigma_{crit}$ and instead involves finding the largest particles in which coherent reversal still dominates.[18] Therapeutic limitations on the allowable $H_0 f$ product[19, 20] and the experimental challenges of producing arbitrarily high fields at high frequencies make it more pragmatic to reframe the problem of optimization as one of adjusting material parameters to make best use of a set of available driving conditions. This observation additionally suggests the possibility of multiplexed heating by selecting materials with properties that lead to dramatically different optimal driving conditions with a similar $H_0 f$ product.

To experimentally test the predictive power of DH simulations, calorimetry measurements at various AMF amplitudes and frequencies were performed on a series of $Fe_3O_4$ SDMNP sets with diameters 10-25nm as determined by transmission electron microscopy (TEM, see Fig. 1 of Ref. 19). SDMNP samples were synthesized and made water soluble as described previously[21] and their concentrations were determined using inductively coupled plasma atomic emission spectroscopy (ICP-AES, ~2 mg Fe/mL for all samples). A vibrating sample magnetometer was used to measure the low field susceptibility of the samples at room temperature, which was then fitted to the Langevin function under the assumption of randomized particle orientation at low fields.[14, 22] As an intuitive proxy for the magnitude of the moment, "magnetic diameter" was defined by considering the necessary diameter of a spherical particle with bulk magnetization to exhibit the same moment.

Calorimetry measurements were conducted with five separate homemade AMF coils driven by a 200W amplifier (Electronics & Innovation, Inc.). The two coils used to produce AMFs with amplitudes exceeding 5kA/m at 100kHz and 500kHz consisted of a ferrite toroid (Ferroxcube, 3F3) machined to include a gap and wrapped with 1050 strand 40 gauge Litz wire (MWS Wire Industries). Calorimetric measurements at lower AMF amplitudes were conducted using simple solenoid coils that did not incorporate soft ferromagnetic components, allowing for operation at higher frequencies. AMF amplitudes were measured with homemade inductive field probes. Thermal effects of resistive power losses in all coils were offset by ice water cooling and the sample was insulated to approach adiabatic conditions. Specific loss power (SLP) values were calculated from linear fits to temperature versus time plots, with small background heating rates measured for water samples subtracted as controls.

AMF amplitudes accessible by coils with soft ferromagnetic cores are limited by the saturation characteristics of the core material. In our setup, operation at 100kHz permitted higher AMF amplitudes to be reached than operation at 500kHz. This allowed the observation of saturation behavior in SLP with increasing AMF amplitude at 100kHz for SDMNPs in or near the ferromagnetic regime (Fig. 2(a)). Though both LRT and DH predict saturation with respect to magnetization, only DH additionally takes into account saturation in the field axis of hysteresis loops (Figs. 2(b) and 2(c)). At 500kHz, driving conditions were limited to lower AMF amplitudes, but sample heating can be confidently measured at smaller amplitudes than 100kHz and compared to predictions from LRT and DH modeling (Figs. 2(d)-2(f)). Functional dependence of SLP on field amplitude can be compared for the smallest and largest SDMNPs. Fig. 2(d) shows that SDMNPs with larger diameter (and thus larger $\sigma$ values) exhibit SLPs that increase more rapidly with increasing AMF amplitude than smaller SDMNPs, suggesting the effect of the applied field on the effective barrier to reversal.

Power law fits can be performed on SLP or specific loss energy per cycle vs. $H_0$ in order to make quantitative comparisons to the predictions of the DH model (Figs. 3(a) and 3(b)). The power law dependence of the largest (most ferromagnetic) SDMNPs has direct relevance to the prospects of magnetothermal multiplexing since the technique relies upon the AMF amplitude dependence of larger particles. In both simulated (Fig. 3(c)) and experimental (Fig. 3(d)) results, the largest SDMNPs demonstrate higher exponents than smaller ones and raising the frequency increases these exponents, particularly for larger particles. Although the DH predicts higher power law dependence for the largest particles than what is observed, this could be attributable to inaccurate assumption of the effective $\sigma$ value, rather than a failure of the model.

A simple proof of concept for multiplexed heating in a single material system is to use $Fe_3O_4$ SDMNPs of different diameters, establishing a low frequency, high amplitude AMF mode (e.g., 100kHz, 35kA/m) and a high frequency, low amplitude AMF mode (e.g., 2.5MHz, 5kA/m) (Fig. 4). The higher amplitude mode favors heat dissipation by larger SDMNPs in the ferromagnetic regime. The lower amplitude mode dramatically reduces the heating of the larger particles, a fact anticipated from the power law dependence of their specific loss power on field amplitude. Increasing the frequency of the lower amplitude mode shifts $\sigma_{crit}$ downward, allowing the smaller SDMNPs to heat more efficiently. This behavior is predicted by DH (Fig. 4, lines) and observed experimentally (Fig. 4, markers). In the low frequency, high AMF amplitude mode, the large 25nm $Fe_3O_4$ SDMNP sample exhibits an SLP 14.9 times higher than the smaller 15 nm $Fe_3O_4$ SDMNP sample slightly doped with Mn (stoichiometric $Mn_xFe_{2-x}O_3$, x = 0.04, as measured by ICP-AES). The high frequency, low AMF amplitude mode reverses the situation, with the smaller SDMNP sample exhibiting an SLP 8.9 times higher than the larger particle sample. Both types of particles reach approximately 400W/g in the respective AMF conditions that favor higher heating.

In biological signaling applications, the local heat dissipation or loss power (LP) of individual SDMNPs rather than the SLP may provide a more relevant metric, particularly under the assumption that targeted cell membranes are coated with a particular surface density of SDMNPs. Although multiplexing SLPs was demonstrated experimentally, the mismatch in individual SDMNP volume between the two types of SDMNPs indicates that multiplexing LPs occurred to a lesser extent, particularly in the higher frequency mode where the 15nm SDMNP LP is estimated to only be about 1.9 times higher than that of the 25nm particle. However, multiplexing LPs should in principle be achievable by selecting two material types with significantly differing $H_k$ values. This approach may have the additional advantage of accomplishing multiplexing with more similar $H_0 f$ products. If field amplitudes of hundreds of kA/m could be achieved at frequencies on the order of 10kHz and a material with a corresponding $H_k$ could undergo coherent reversal, the same strategy could be extended to three or more modes. Controlling composition over the wide range of magnetocrystalline anisotropy values for $A_xFe_{3-x}O_4$ (A = Mn, Fe, Co) or creating exchange coupled core shell architectures could provide the means to produce particle sets with $\sigma$ and $H_k$ values tailored to multiplexing. $\sigma$-$\xi$ space provides a graphical representation onto which either approach could be equivalently mapped and understood.

We experimentally observed predictive successes of DH theory by examining detailed calorimetric data for a range of $Fe_3O_4$ SDMNP sizes driven by a variety of AMF amplitudes and frequencies. By developing a graphical representation to more intuitively understand DH theory's predictions and refining the question of materials optimization to take into account driving conditions, the possibility of multiplexed magnetic heating has been anticipated and experimentally demonstrated. This physical effect may allow multiple biological or biochemical processes to be addressed via heat mediated signaling schemes, even when the targeted actions occur in close proximity.

**Acknowledgements:**

This work was funded in part by the Sanofi Biomedical Innovation award and the Defense Advanced Research Project Agency Young Faculty Award to P.A. This work made use of the MRSEC Shared Experimental Facilities supported by the National Science Foundation (NSF) under award number DMR - 0819762. M.G.C. is supported by the National Defense Science and Engineering Graduate Fellowship Program and R.C. is supported by the NSF Graduate Fellowship Program. We extend our gratitude to C.A. Ross for insightful comments on our manuscript.

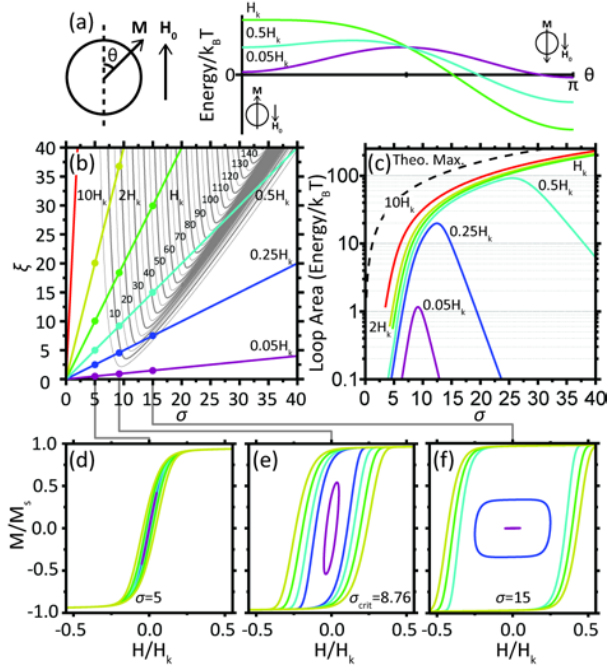

FIG. 1 (Color online) Graphical representation of $\sigma$-$\xi$ space. (a) A schematic of the SDMNP in the AMF and energy as a function of θ (angle between the magnetic moment and the easy axis of the SDMNP). (b) Contour plot of hysteresis loop area as a function of $\sigma$ and $\xi$ for uniaxial anisotropy at 500kHz, superimposed with paths representing AMF amplitudes of different magnitude relative to $H_k$. (c) Hysteresis loop area as a function of $\sigma$ plotted along the paths in (b). (d)-(f) Simulated hysteresis loops for points in $\sigma$-$\xi$ space for a representative $\sigma$ values from the superparamgentic regime (d), ferromagnetic regime (f), and the $\sigma_{crit}$ dividing them (e), which varies with frequency.

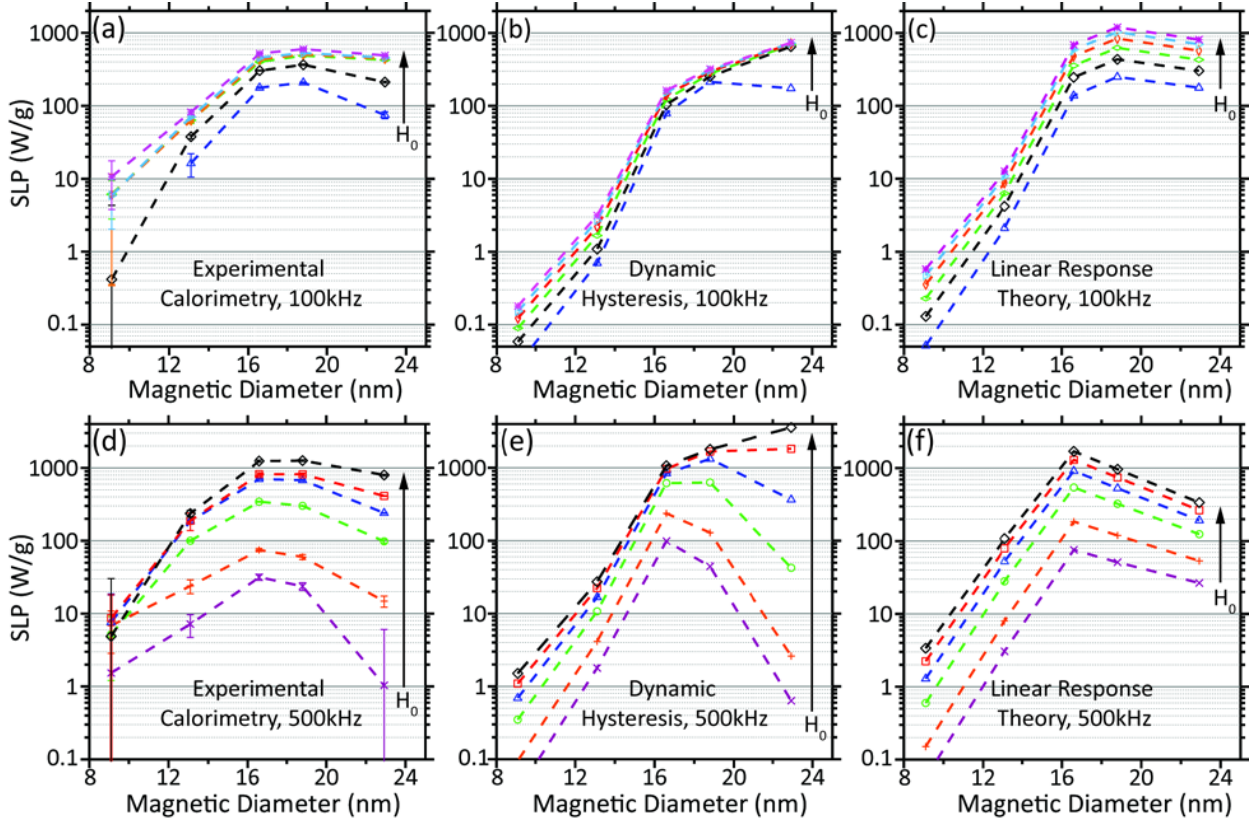

FIG. 2 (color online) Experimental calorimetry data for SLP vs. magnetic diameter with increasing field is compared against the predictions of DH and LRT simulations. (a) SLPs at AMF of frequency of 100kHz and amplitudes 15-65 kA/m measured for $Fe_3O_4$ SDMNPs in aqueous solutions (~2mg Fe/mL), with vertical error bars representing the standard deviation over five trials. (b), (c) SLPs simulated by DH (b) and LRT (c) at AMF parameters identical to the experiment in (a). (d) SLPs at AMF of frequency of 500kHz and amplitudes 3-25 kA/m measured for $Fe_3O_4$ SDMNPs as described in (a). (e), (f) SLPs simulated by DH (e) and LRT (f) at AMF parameters identical to the experiment in (d). Different markers denote the AMF amplitudes: ✕ 3kA/m, ✚ 5kA/m, ◯ 10kA/m, △ 15kA/m, ☐ 20kA/m, ◇ 25kA/m, ⬖ 35kA/m, ◊ 45kA/m, ▷ 55kA/m, and ✳ 65kA/m (all ±<3% in experiments (a) and (d)).

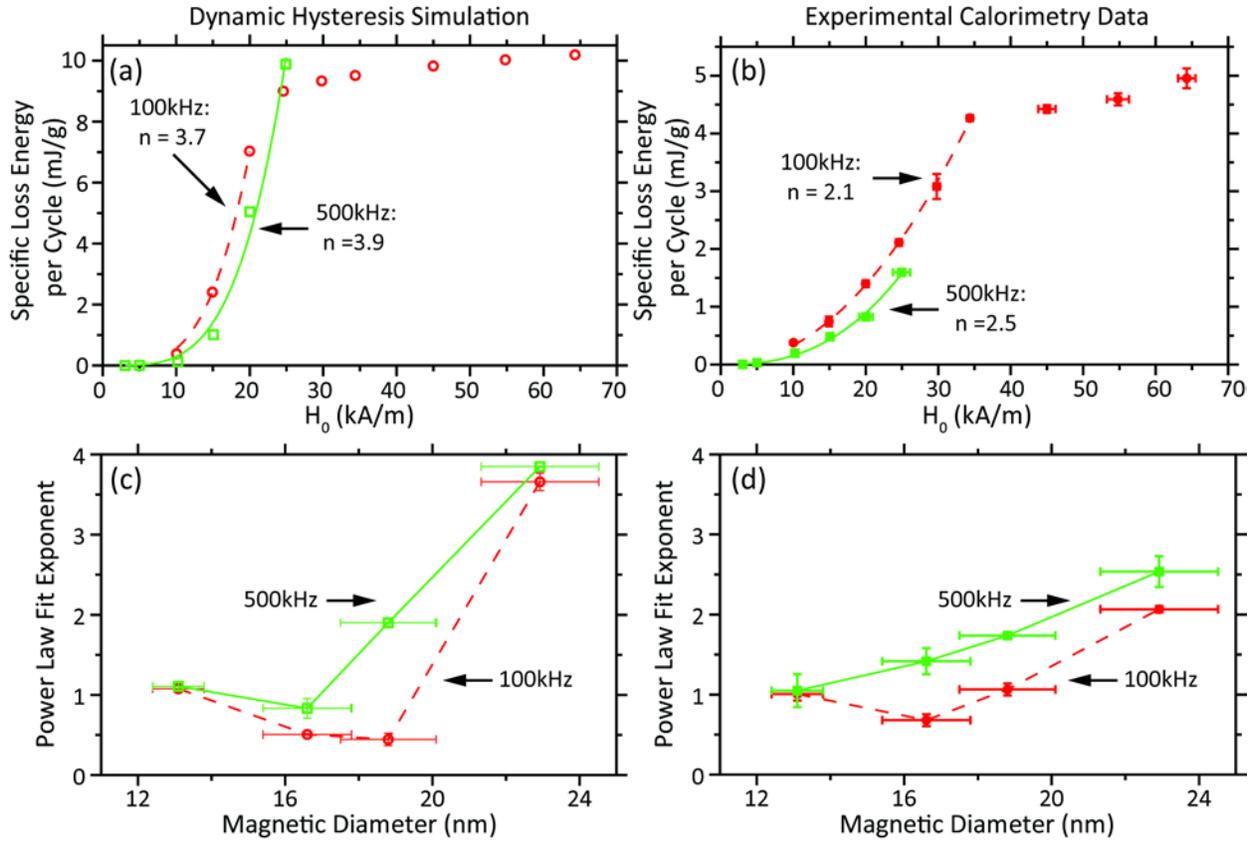

FIG. 3 (color online). (a) Power law fits (lines) of specific loss energy per cycle as a function of field amplitude $H_0$ simulated data (open markers, ○ 100kHz, □ 500kHz ) and (b) experimental calorimetry (solid markers, ● 100kHz, ■ 500kHz ) are shown for the $d_m$=23 ± 1.6 nm $Fe_3O_4$ SDMNP set. The exponents extracted from identical fits on SDMNP sets of various diameters are shown for (c) simulated and (d) experimental data as a function of SDMNP size. In all panels: 100kHz dashed line, 500 kHz solid line.

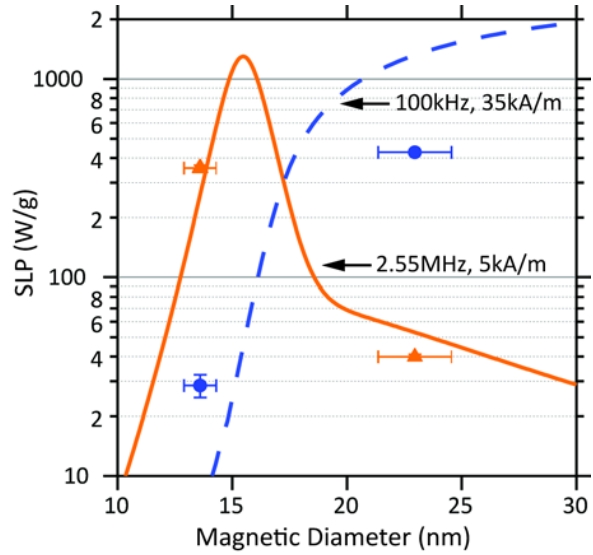

FIG 4. (color online) SLP multiplexing results are shown for a $d_m$=23nm $Fe_3O_4$ particle and a 15nm $Mn_xFe_{3-x}O_4$ (x=0.04) particle. For a 100kHz, 35kA/m AMF (● experimental, ── ── dynamic hysteresis simulation) the larger SDMNP in the ferromagnetic regime heats more dramatically (SLP = 426±5 W/g) than the smaller SDMNP in the superparamagnetic regime (SLP = 29±4 W/g). Lowering the amplitude to 5kA/m and raising the frequency to 2.55MHz (▲ experimental, ──── linear response theory approximation including Brownian relaxation) swaps the heat dissipation rates of the two SDMNP sets (larger SDMNP SLP =40.±1 W/g, and smaller SDMNP SLP = 356±12 W/g).